\documentclass[trackchanges]{aastex7}
\usepackage{CJK}
\usepackage{subcaption}

\begin{document}

\begin{CJK*}{UTF8}{gbsn}  

\title{The Maximum Particle Energy Gain During Magnetic Reconnection}

\author[orcid=0009-0009-7643-6103,sname='Yin']{Zhiyu Yin (尹志宇)}
\affiliation{Department of Physics, University of Maryland, College Park, MD 20742, USA}
\affiliation{IREAP, University of Maryland, College Park, MD 20742, USA}
\email[show]{zyin1@umd.edu} 

\author[orcid=0000-0002-0449-1498,sname='Arnold']{H. Arnold}
\affiliation{The Johns Hopkins University Applied Physics Laboratory, Laurel, MD, USA}
\email{harry.arnold@jhuapl.edu} 
\author[orcid=0000-0002-9150-1841,sname='Drake']{J. F. Drake}
\affiliation{Department of Physics, University of Maryland, College Park, MD 20742, USA}
\affiliation{IREAP, University of Maryland, College Park, MD 20742, USA}
\affiliation{Institute for Physical Science and Technology, University of Maryland, College Park, MD 20742, USA}
\affiliation{Joint Space Science Institute, University of Maryland, College Park, MD 20742, USA}
\email{drake@umd.edu} 

\author[orcid=0000-0002-5435-3544,sname='Swisdak']{M. Swisdak}
\affiliation{IREAP, University of Maryland, College Park, MD 20742, USA}
\affiliation{Joint Space Science Institute, University of Maryland, College Park, MD 20742, USA}
\email{swisdak@umd.edu}

\begin{abstract}

The factors that control the maximum energy attained by protons and electrons during magnetic reconnection are investigated analytically and using large-scale simulations with the \textit{kglobal} model. Previous work revealed that a strong ambient guide field strongly impacts particle energy gain during reconnection, suppressing energy gain from Fermi reflection by increasing the radius of curvature of reconnected field lines. However, previous simulations have also shown that the maximum energy gain increases with the system size. The physical basis for this result has not been explored. We perform simulations that vary the effective system size over a large range to isolate the processes determining the maximum energy gain. The maximum energy $W_{max}$ is regulated by the number of magnetic-island mergers that occur, as multiple flux ropes that form at early time repeatedly merge until the largest approaches the system scale. Fermi reflection in these repeated mergers dominates particle energy gain. The number of mergers is linked to the effective system size -- larger systems produce a larger number of flux ropes and more mergers. That $W_{max}$ is linked to the number of flux rope mergers has implications for understanding why particle-in-cell simulations only produce powerlaw distributions of energetic particles with a limited range in energy.

\end{abstract}




\section{Introduction} 
\label{sec:intro}
Magnetic reconnection rapidly converts magnetic energy into particle energy in a wide range of environments, including solar flares \citep{lin71,Masuda94,Lin03,Benz17}, Earth's magnetosphere \citep{Dungey61,Sonnerup81}, and the solar wind \citep{Gosling05,Phan06,phan21}. This energy release drives both substantial bulk heating that scales with \( m_i C_A^2 \) \citep{Phan13a,Phan14,Oieroset23,Oieroset24,Oka25}, where $C_A$ is the Alfv\'en speed, and non-thermal particles with power-law energy distributions \citep{Lin03,Oieroset02,Krucker10,Gary18,Ergun20b,Desai25}. {\it In~situ} measurements show that protons typically receive more bulk energy than electrons, and magnetotail observations suggest that protons may also dominate the non-thermal energy budget \citep{Ergun20,Rajhans25}. For solar flares, the proton energy content remains uncertain because measurements rarely extend below \(\sim 1\ \mathrm{MeV}\) \citep{Lin03,Emslie12}, although recent line-broadening observations indicate that ion heating may exceed electron heating \citep{Russell25}.

Reconnection generates bent magnetic field lines whose tension drives an Alfv\'enic exhaust \citep{Parker57,Petschek64,Sato79}, transporting energy away from the x-line and converting magnetic energy into plasma energy \citep{Lin93}. In many systems, the reconnecting current sheet fragments into multiple x-lines and associated flux ropes \citep{Biskamp86,Drake06,Loureiro07,Bhattacharjee09,Daughton11}, consistent with spacecraft observations \citep{Chen08,Phan24}. Particle energization in this environment is dominated by Fermi reflection in growing and merging flux ropes \citep{Drake06,Oka10,Dahlin14,Guo14,Li19,Zhang21}, a process that naturally produces the observed power-law energy distributions in both electrons \citep{Arnold21} and ions \citep{Zhang21,Yin24b}. Because the Fermi energy gain rate scales with the energy of a particle \citep{Drake06a,Drake13}, the highest-energy particles are accelerated most efficiently, leading to extended non-thermal tails. Thus, the energy released during reconnection is not distributed evenly among accelerated particles. Recent {\it in situ} observations of a reconnection event in the near-solar heliospheric current sheet supported this picture. The observations revealed that the maximum proton energy was around 500 keV or $10^3$ times the available magnetic energy per particle $m_iC_A^2\sim 0.5$ keV. 

The \textit{kglobal} model \citep{Arnold19,Drake19,Yin24}, which removes kinetic scales to describe reconnection in macroscopic systems, was the first fully self-consistent framework to produce the extended power-law spectra seen in observations. Simulations using its upgrade to include particle protons \citep{Yin24} demonstrated that energetic protons gain substantially more energy than electrons even when both species begin with equal temperatures \citep{Yin24b}. Further simulations revealed that the preferential proton energization resulted from the significant energy gain of ions on their first entry into reconnection exhausts \citep{Yin25}. Because of this headstart electron energy gain never catches that of the ions. However, an important open question has remained: what sets the upper limit on the energy that a single particle can attain during reconnection? Specifically, while the dependence of particle energization on the ambient guide field is well understood \citep{Drake06a,Drake13,Dahlin16}, the dependence of the maximum energy on the system size is not \citep{Arnold21,Yin24b}. The present work addresses this question directly by identifying the mechanism and scaling that control the dependence of the maximum particle energy during reconnection on the system size.

In this paper, we use \textit{kglobal} simulations to investigate the heating and energization of protons and electrons during magnetic reconnection and to identify the physical processes responsible for controlling the maximum energy that particles gain. Section~\ref{sec:theory} outlines the theoretical framework relevant for understanding particle energization in macroscale reconnection. Section~\ref{sec:modelsetup} describes the simulation setup, including normalization, parameters, and numerical implementation. Section~\ref{sec:results} presents the primary results, focusing on the mechanisms that control the maximum energies of the two species. Finally, Section~\ref{sec:conclusion} summarizes the main conclusions and discusses the implications of these findings for reconnection-driven particle acceleration in the heliosphere and astrophysical systems.

\section{Theoretical Background}
\label{sec:theory}
What physics limits the maximum energy gain of particles undergoing acceleration during reconnection remains an important question. Models, including PIC, hybrid and {\it kglobal} have produced data related to this question. In the case of PIC and hybrid simulations, particles with the largest energy can become demagnetized, inhibiting further energy gain and therefore producing an upper limit on particle energy. However, because of the inadequate separation of kinetic from macro-scales in PIC and hybrid models, the demagnetization upper limit is unlikely to be relevant to most real reconnecting systems. For example, direct measurements in HCS reconnecting events with energetic particles reveal that the widths of reconnection exhausts are thousands of ion inertial lengths \citep{Desai25}.  In the case of {\it kglobal} simulations, the Larmor radii of the particles are ordered out of the equations so there is no demagnetization limit. The system is normalized to an arbitrary length $L$ and times are normalized to the Alfv\'en transit time. As a result, the physical scale length $L$ enters only through the dissipation parameter that facilitates reconnection, which is taken to be a hyper-resistivity $\nu$. (A normal resistivity is overly dissipative.) The effective system size is therefore controlled by an effective Lundquist-like number 
\begin{equation}
    S_\nu=\frac{C_AL^3}{\nu}.
    \label{eq:S_nu}
\end{equation}
Simulations have revealed that the upper limit on particle energy does depend on $S_\nu$. This is reflected in an upper limit on the powerlaw distribution of the most energetic particles \citep{Arnold21,Yin24b} (see also Fig.~\ref{fig:combined_spectra} below). Larger domains, corresponding to larger values of $S_\nu$, produce particles with greater energy. The physics behind this result has not been previously explored.

We hypothesize that the peak particle energy achieved during reconnection is governed by the number of magnetic island mergers that occur during the system's evolution. The energy gain of particles during island merger has been explored in simulations \citep{Oka10,Drake10} and analytically \citep{Drake13}. The merger of two equal-sized flux ropes approximately doubles the energy of all particles within the flux ropes. This result is a consequence of the field line shortening that takes place during reconnection and the conservation of the second adiabatic invariant $v_\parallel s$, where $v_\parallel$ is the magnetic field aligned velocity and $s$ is the length of the field line. Consider a magnetic field line that wraps two adjacent flux ropes each with radius $r_0$. The length of this field line is $s_0=4\pi r_0$. After the flux ropes undergo a nearly incompressible merger, the resultant flux rope has a radius $\sqrt{2}r_0$ and a field line length $s=2\sqrt{2}\pi r_0=s_0/\sqrt{2}$. The particle velocity therefore increases by $\sqrt{2}$ and the energy by a factor of 2. Thus, the energy gain of a particle that experiences $N$ mergers with equal sized flux ropes should be given by 
\begin{equation}
    W=W_i2^N
\end{equation}
where $W_i$ is the initial particle energy. To test this idea, we need to estimate the number of mergers in a system of a given size. The merging process starts with the smallest flux ropes, with merging continuing until at late time a single system-scale flux rope remains. Thus, to determine the number of mergers we need to estimate the scale size of the smallest scale islands that form early in the simulation. 

To determine the width $w_0$ of the smallest islands, we estimate the current layer thickness in the presence of a fourth-order hyper-resistivity $\nu$. The evolution equation for magnetic flux $\psi$ is given by 
\begin{equation}
    \frac{\partial \psi}{\partial t} + \mathbf{v}\cdot\nabla \psi = \nu \nabla^4 \psi.
\end{equation}
Balancing the convective inflow of magnetic flux with with the hyper-resistivity term gives
\begin{equation}
    \frac{v_{\rm rec}}{w_0} \sim \nu \frac{1}{w_0^4},
\end{equation}
where $v_{\rm rec}$ is the inflow velocity into a reconnection site. Thus,
\begin{equation}
    v_{\rm rec} \sim \frac{\nu}{w_0^3}.
\end{equation}
Balancing the inflow of plasma with the outflow while ignoring compression, we obtain

\begin{equation}
    v_{\rm rec} L \sim C_A w_0,
\end{equation}
where $L$ is the system size and $C_A$ is the  Alfv\'en speed. Eliminating $v_{\rm rec}$ yields an expression for $w_0$
\begin{equation}
    w_0 \sim L S_\nu^{-1/4}.
    \label{eq:w0}
\end{equation}
In linear tearing mode theory, the strongest growing mode has a scale length of order of $w_0$, so $w_0$ in Eq.~(\ref{eq:w0}) is the minimum size of flux ropes expected in the system. 

A single merger increases the island area by a factor of $\sim 2$, so each island merger roughly increases the island size by the factor $\sqrt{2}$. 
After $N$ mergers, the island radius becomes
\begin{equation}\label{eq:radius_ratio}
    w_N \sim 2^{N/2} w_0.
\end{equation}
When the island grows to the system scale, $w_N \sim L$, 
\begin{equation}
    2^{N/2} \sim \frac{L}{w_0},
\end{equation}
so
\begin{equation}
    2^{N} \sim \left( \frac{L}{w_0} \right)^2\sim S_\nu^{1/2}.
\end{equation}
Thus, the maximum energy $W_{max}$ of particles in a {\it kglobal} simulation should scale as 
\begin{equation}
    W_{\max} \sim W_i2^N\sim W_i \left( \frac{L}{w_0} \right)^2 \sim W_i S_\nu^{1/2}.
    \label{eq:Wmax}
\end{equation}
The maximum particle energy during reconnection in the \textit{kglobal} model should increase with the effective system size and the maximum energy should be linked to the number of island mergers, which is controlled by $S_\nu$. These ideas need to be benchmarked with simulation data.

\section{Simulation Model Setup}
\label{sec:modelsetup}
To examine how the maximum particle energy depends on the system size, we carry out a series of simulations spanning effective Lundquist numbers from $S_{\nu}=1.2\times10^{7}$ to $6.1\times10^{9}$, as shown in Table \ref{tab:initial_setting}. The simulations employ the \textit{kglobal} model \citep{Yin24b}, which evolves an MHD fluid composed of proton and electron species together with macro-particle protons and electrons advanced in the guiding-center limit. This model removes the need to resolve kinetic scales such as Larmor radii while still capturing particle energization. Particles drift across the magnetic field via $\mathbf{E}\times\mathbf{B}$ motion while streaming along the magnetic field at their parallel velocities.

In the standard \textit{kglobal} normalization \citep{Yin24b}, the upstream reconnecting magnetic field \( B_0 \) and total proton density \( n_{i0} \) define the  Alfv\'en speed
\[
    C_{A0} = \frac{B_0}{\sqrt{4\pi m_i n_{i0}}}.
\]
Lengths are normalized to a macroscopic scale \( L_0 \), times to the Alfv\'en time \( \tau_A = L_0 / C_{A0} \), and particle energies to \( m_i C_{A0}^2 \). The perpendicular electric field scales as \( C_{A0} B_0 / c \), while the parallel electric field follows from electron force balance and is given by \( m_i C_{A0}^2 / (e L_0) \); although small in magnitude, the associated potential drop over the scale \( L_0 \) remains comparable to the magnetic energy per proton. A reduced proton--electron mass ratio of \(25\) is used, but the results presented here are insensitive to this choice \citep{Yin25}.

The speed of light is set to \(30\,C_{A0}\) in all simulations. All species (fluid and particle) begin with an isotropic temperature of \(0.0625\, m_i C_{A0}^2\), with the particles initially given Maxwellian velocity distributions. The system is initialized as a force-free current sheet with uniform densities and pressures and periodic boundary conditions. The magnetic field is given by
\[
\mathbf{B} = B_0 \tanh\!\left(\frac{y}{w}\right)\mathbf{\hat{x}}
    + \sqrt{B_0^2\, \text{sech}^2\!\left(\frac{y}{w}\right) + B_g^2}\,\mathbf{\hat{z}},
\]
where \(B_g\) is the guide field and the current-sheet half-width is \(w = 0.005\,L_0\). The total electron (\(n_e\)) and proton (\(n_i\)) densities are normalized to unity, with particles comprising 25\% of each species (\(n_{ep}\), \(n_{ip}\)) and the remaining 75\% treated as a fluid (\(n_{ef}\), \(n_{if}\)). The results are insensitive to this choice of particle fraction \citep{Yin24b}.

Simulations are performed on grids ranging from \(1024\times512\) to \(8192\times4096\) cells, as listed in Table~\ref{tab:initial_setting}, with 100 particles per cell. The timestep \(\delta t\) is varied with resolution to maintain accurate particle advancement (e.g., \(c\,\delta t \simeq \delta\), where \(\delta\) is the grid spacing), and the corresponding values of \(\delta t\) are reported in Table~\ref{tab:initial_setting}.

\begin{deluxetable*}{ccccc}
\tablewidth{0pt}
\tablecaption{Parameters for Simulation Domains\label{tab:initial_setting}}
\tablehead{
  \colhead{Grid Points} & 
  \colhead{$1024 \times 512$} &
  \colhead{$2048 \times 1024$} & 
  \colhead{$4096 \times 2048$} & 
  \colhead{$8192 \times 4096$}
}
\startdata
Time Step (in $\tau_A$) & $2 \times 10^{-4}$ & $1 \times 10^{-4}$ & $5 \times 10^{-5}$ & $2.5 \times 10^{-5}$ \\
Proton Number Density Diffusion ($D_n$) & $10.5 \times 10^{-5}$ & $5.25 \times 10^{-5}$ & $2.625 \times 10^{-5}$ & $1.3125 \times 10^{-5}$ \\
Proton Pressure Diffusion ($D_p$) & $10.5 \times 10^{-4}$ & $5.25 \times 10^{-4}$ & $2.625 \times 10^{-4}$ & $1.3125 \times 10^{-4}$ \\
Hyperviscosity ($\nu_B$, $\nu_{nv}$, $\nu_n$, and $\nu_p$) & $84 \times 10^{-9}$ & $10.5 \times 10^{-9}$ & $1.312 \times 10^{-9}$ & $0.1640 \times 10^{-9}$ \\
Effective Lundquist Number ($S_\nu$) & $1.2 \times 10^{7}$ & $9.5 \times 10^{7}$ & $7.6 \times 10^{8}$ & $6.1 \times 10^{9}$ \\
\enddata
\end{deluxetable*}

In our simulations, explicit diffusion and hyperviscosity terms are incorporated to maintain numerical stability and to suppress high-frequency noise at the grid scale. The diffusion coefficients for number density (\(D_n\)) and pressure (\(D_p\)) are assigned the values listed in Table~\ref{tab:initial_setting}.  A fourth-order hyperviscosity (\(\nu\)) is included in the magnetic-field evolution to enable reconnection while minimizing large-scale dissipation. This term enters the induction equation as a \(\nabla^4\) operator, and the same form of hyperviscosity is applied to the evolution of the fluid proton flux, number density, and pressure. The corresponding effective Lundquist number (see Equation (\ref{eq:S_nu}))
controls the ratio of macroscopic to dissipation scales and is varied to adjust the effective system size.  The hyperviscosity coefficients (\(\nu_B\), \(\nu_{nv}\), \(\nu_n\), and \(\nu_p\)) and the resulting values of \(S_\nu\) used in each run are summarized in Table~\ref{tab:initial_setting}.

\section{Simulation Results} 
\label{sec:results}

To quantify how the system size influences magnetic-island formation and the resulting particle energization, we evaluate the island characteristics in each simulation and summarize the results in  Table~\ref{tab:island_mergers}.  
The initial island size is the characteristic width along $x$ of the smallest islands that appear at the onset of reconnection (around $t=1\tau_A$).
To calculate this flux rope size, we count the number of $B_y=0$ crossings in a cut along the center of the current sheet. The number of flux ropes is half of the number of zero crossings. The early time island size then follows by dividing the domain size along $x$ by the number of flux ropes. The final flux rope size corresponds to the $x$-axis extent of the dominant system-scale flux rope that remains after successive mergers. The ratio of the final to the initial island size therefore reflects the total geometric growth of the island hierarchy. Using this ratio and Eq.~(\ref{eq:radius_ratio}), we infer the number of island mergers, $N$, which increases systematically with system size. These quantities establish the connections between domain size, the number of mergers, and the expected maximum particle energy, and provide the basis for interpreting the spectra.

\begin{deluxetable*}{ccccc}
\tablewidth{0pt}
\tablecaption{Island Structure and Merger Parameters for Different Simulation Domain Sizes\label{tab:island_mergers}}
\tablehead{
  \colhead{Box Size (grid points)} & 
  \colhead{$1024 \times 512$} &
  \colhead{$2048 \times 1024$} & 
  \colhead{$4096 \times 2048$} & 
  \colhead{$8192 \times 4096$}
}
\startdata
Number of $B_y$ zero crossings at $t=1\tau_A$     & 56  & 64   & 98   & 154  \\
Initial island size (grid points)   & 36.57 & 64 & 83.6 & 106.39 \\
Final big island size (grid points) & 574 & 1100 & 2200 & 3000 \\
Ratio of final to initial island size & 15.7 & 17.19 & 26.3 & 28.20 \\
Number of mergers                   & 7.95 & 8.21 & 9.43 & 9.64 \\
\enddata
\end{deluxetable*}

\begin{figure*}[ht!]
\centering
\includegraphics[width=0.9\textwidth]{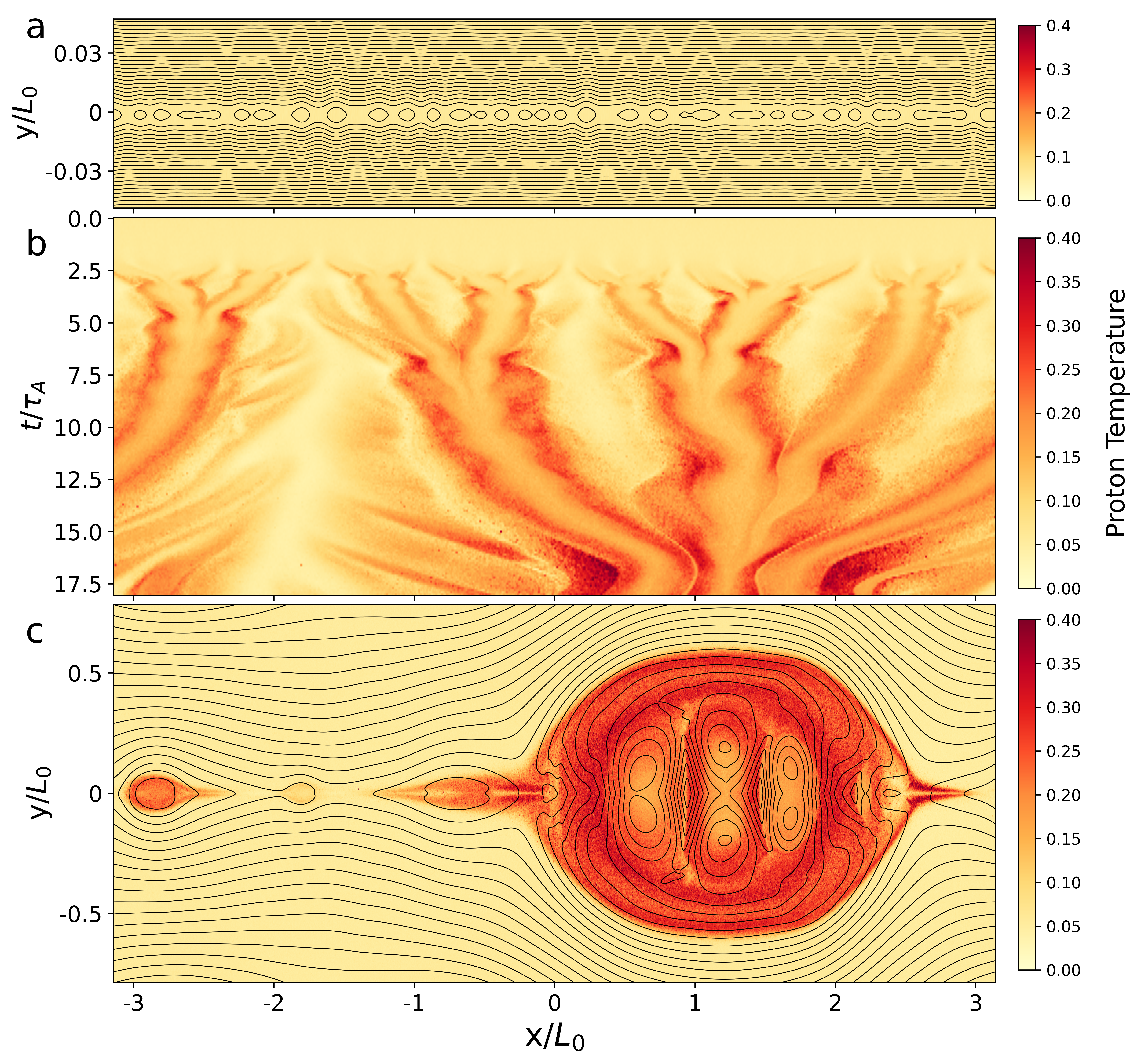}
\caption{Proton temperature evolution in a reconnecting current sheet.
(a) Early-time two-dimensional proton temperature at 
$t \simeq 1\,\tau_A$, together with magnetic-flux contours, 
showing the onset of reconnection along the current sheet and the
formation of multiple small magnetic islands.
(b) Space--time evolution of the proton temperature along the center of the current sheet.
The formation, growth, and merging of magnetic islands produce
filamentary heating structures and progressively higher temperatures
at later times.
(c) Late-time proton temperature with magnetic-flux contours.
A large system-scale island has formed through successive mergers.
}
\label{fig:combined_spacetime}
\end{figure*}

To illustrate the evolution of island formation and plasma particle heating within a single simulation, we present in Figure~\ref{fig:combined_spacetime} proton temperature diagnostics for the second-smallest simulation domain ($2048\times 1024$). Panel (a) displays the early-time state at $t\simeq 1\tau_A$, where reconnection has already begun and multiple small islands have formed along the current sheet. Panel (b) displays the space–time evolution of the proton temperature along the center of the current sheet, revealing the sequential growth and coalescence of flux ropes and the associated proton heating. By the late stage, shown in panel (c), successive mergers have produced a large system-scale island whose interior contains the largest temperature enhancement. The progressive increase in island size with time illustrates the hierarchical dynamics described in Sec.~\ref{sec:theory} and summarized in Table~\ref{tab:island_mergers}, and provides a direct visual demonstration of the merger-driven heating process that ultimately shapes the particle energy spectra.

\begin{figure*}[ht!]
\centering
\includegraphics[width=0.9\textwidth]{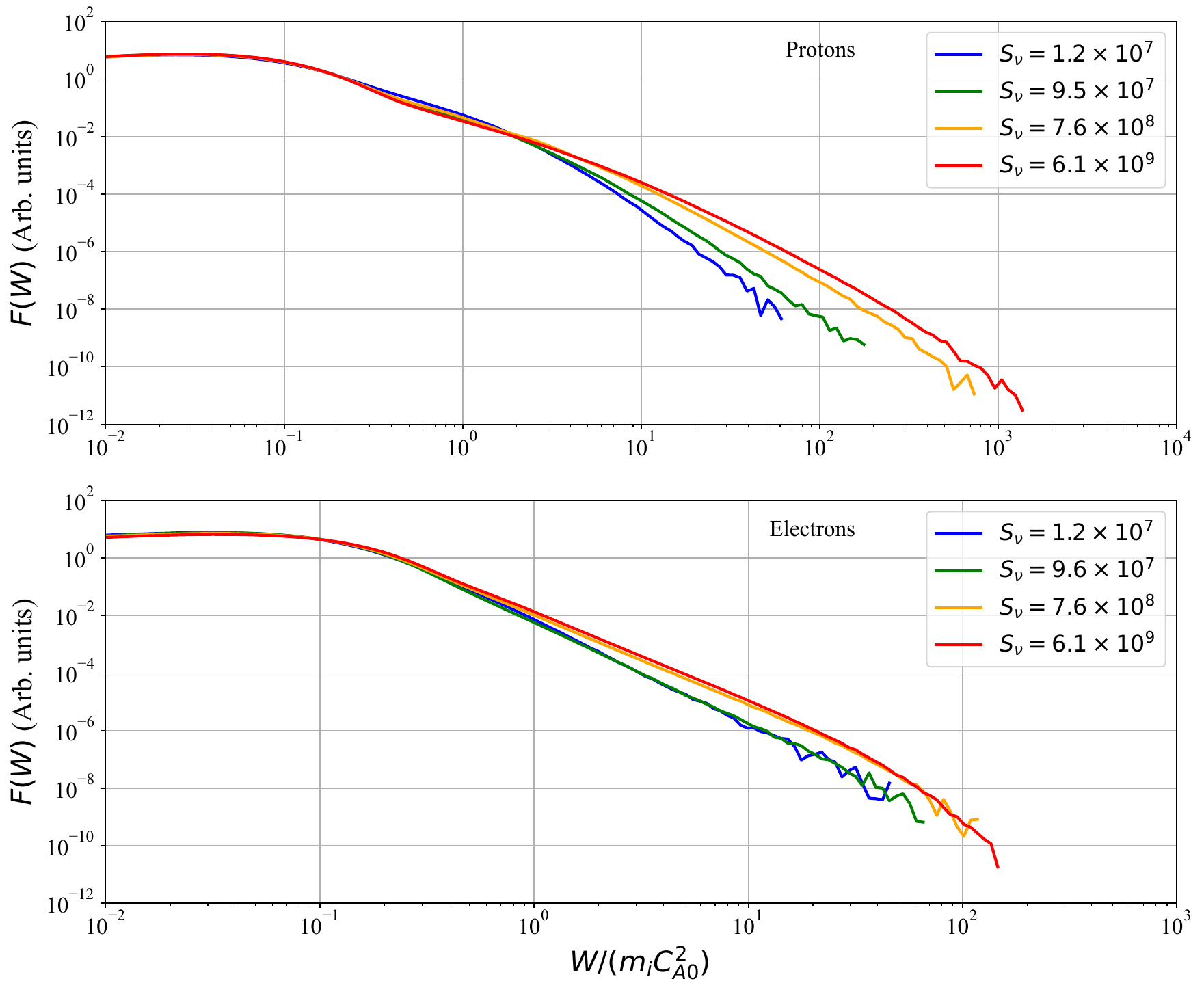}
\caption{Proton (top) and electron (bottom) energy spectra for different effective system sizes $S_{\nu}$, shown using the standard normalized energy $W/(m_i C_{A0}^2)$. 
The high-energy cutoff increases systematically with $S_{\nu}$, reflecting the stronger particle acceleration that takes place in larger simulation domains.
}
\label{fig:combined_spectra}
\end{figure*}

The impact of system size on the resulting particle spectra is shown in Figure~\ref{fig:combined_spectra} in the late-time proton and electron distributions for the four values of $S_{\nu}$. For both species, the spectra exhibit a thermal core and an extended power-law tail whose slope varies only modestly with system size once the domain becomes sufficiently large. As $S_{\nu}$ increases, the high-energy cutoff shifts to progressively larger energies. This behavior reflects the greater number of flux ropes that form in larger systems and the larger number of mergers available to accelerate particles. In these larger domains, particles undergo more Fermi reflections before the system reaches its final system-scale island, consistent with the increasing merger counts listed in Table~\ref{tab:island_mergers}. These additional mergers lead to enhanced energization and higher maximum energies.

\begin{figure*}[ht!]
\centering
\includegraphics[width=0.9\textwidth]{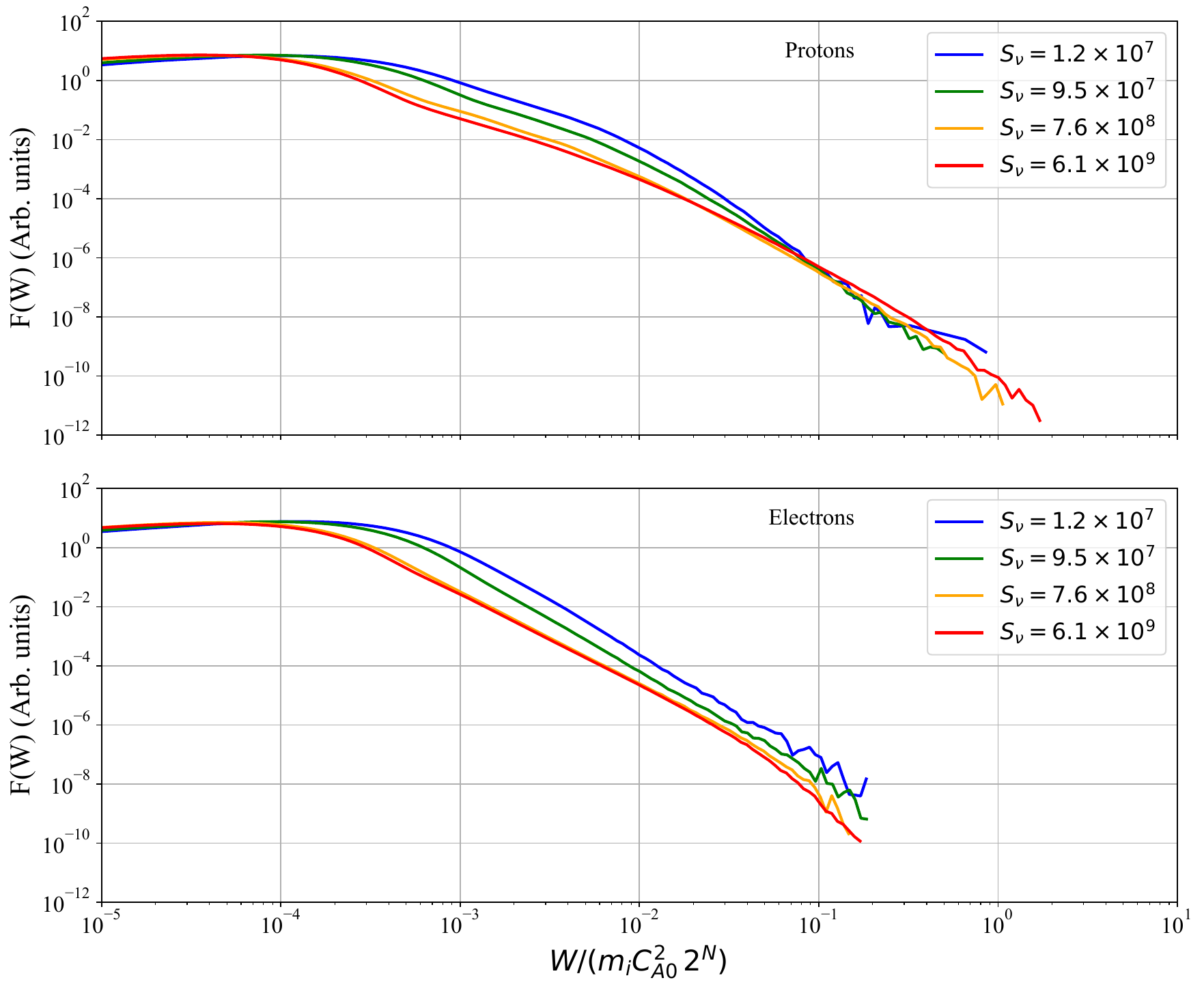}
\caption{Proton (top) and electron (bottom) energy spectra for different effective system sizes $S_{\nu}$, shown using the merger-scaled energy normalization $W/(m_i C_{A0}^2 2^{N})$. 
Here $N$ denotes the number of island mergers characteristic of each system size. Division by $2^{N}$ collapses the upper energy limit to a common initial energy (see Eq.~(\ref{eq:Wmax})). 
This scaling reduces systems with differing effective sizes to a common normalized energy frame.}
\label{fig:combined_normalized_spectra}
\end{figure*}

To test the theoretical prediction that the maximum energy scales with the number of mergers, we plot the same spectra using the merger-scaled normalization $W/(m_i C_{A0}^{2} 2^{N})$, where $N$ is the number of mergers extracted from Table~\ref{tab:island_mergers}. As shown in Figure~\ref{fig:combined_normalized_spectra}, this scaling collapses the high-energy ends of the spectra for all system sizes onto curves with a common upper endpoint. This endpoint corresponds to the starting energy of the two species since, from Eq.~(\ref{eq:Wmax}), $W_i\sim W_{max}/2^N$. An important result is that the starting energies of the protons and electrons are not equal with $W_i\sim m_iC_A^2$ for protons and $W_i\sim 0.2m_iC_A^2$ for electrons. This result is perhaps surprising since the initial temperature of both species is $0.0625m_iC_A^2$. The reason for this difference is that protons gain an energy that scales as $m_iC_A^2$ on a single entry into a reconnection exhaust, while electrons, because of their small mass, gain only a small fraction of $m_iC_A^2$ on a single exhaust entry \citep{Yin25}. Thus, the starting energy $W_i$ of electrons reflects their initial ambient energy while the starting energy of the protons reflects their first entry into a reconnection exhaust. It was shown previously that if the initial energy of electrons is increased sufficiently above that of the protons, the total electron energy gain can exceed that of the protons \citep{Yin25}. 

These results provide clear numerical support for the proposition that particles gain energy through repeated Fermi reflections as islands grow and merge. Since the number of mergers increases as $N \sim \ln(\sqrt{S_{\nu}})$, the maximum energy follows the predicted scaling $\overline{W}_{\max} \propto S_{\nu}^{1/2}$. The simulations thus confirm that the dominant control parameter for the highest-energy particles, for a fixed value of the ambient guide field, is the smallest island scale set by hyperviscous dissipation, which in turn determines how many mergers occur before the system reaches a single dominant flux rope. This merger-based picture explains both the system-size dependence of the spectral cutoff and the consistent proton–electron differences observed across all runs -- the effective starting energy of electrons is smaller than that of the protons \citep{Yin25}. 

\section{Conclusion} \label{sec:conclusion}

We have used large-scale \textit{kglobal} simulations to identify the mechanisms that determine the maximum energy acquired by protons and electrons during magnetic reconnection in macroscale current sheets. By spanning more than two orders of magnitude in the effective Lundquist number, $S_\nu=L^3C_A/\nu$ with $\nu$ the hyper-resistivity, the simulations isolate the role of magnetic-island growth and merging in controlling particle energization. The island characteristics reveal that larger domains generate more flux ropes and undergo more mergers, consistent with the multi-x-line reconnection picture established in kinetic and MHD studies \citep{Biskamp86,Drake06,Loureiro07,Bhattacharjee09,Daughton11}. The maximum particle energy $W_{max}$ scales as $W_i2^N$, where $N$ is the number of flux rope mergers possible in a simulation of a given size and $W_i$ is the initialization energy. A surprise is that the initialization energy differs for electrons and protons even if the two species begin with a common temperature. For protons $W_i\sim m_iC_A^2$ since protons gain an Alfv\'enic kick in velocity on a single entry into a reconnection exhaust \citep{Hoshino98,Gosling05,Yin25}. In contrast, electrons, because of their small mass, gain only a very small energy in a single entry into a reconnection exhaust. Thus, their starting energy $W_i$ is well below that of the protons. This is the reason that their overall energy gain remains well below that of the protons \citep{Yin25}. 

The data on the particle spectra support the idea that protons and electrons have differing effective starting energies and peak energies that are given by $W_{max}\sim W_i2^N$. When plotted versus $W/(m_iC_A^22^N)$,  with the value of $N$ determined by analyzing the data from each simulation (see Table \ref{tab:island_mergers}), the energy spectra display a common upper limit, the initiation energy $W_i$, which is around $m_iC_A^2$ for protons and $0.2m_iC_A^2$ for electrons. The fact that the spectra,  when plotted against the merger normalized energy, display a common upper limit confirms both that protons and electrons have distinct initiation energies and also that the maximum energy gain scales with the number of mergers as $2^N$. This also directly supports the theoretical prediction that the maximum energy follows the scaling
\begin{equation}
W_{\max} \propto S_{\nu}^{1/2},
\end{equation}
since $N=\ln S_\nu/2\ln 2 $. This result provides a quantitative upper bound for particle acceleration in macroscale reconnecting systems.

Protons reach higher maximum energies than electrons across all simulations. This outcome is consistent with earlier \textit{kglobal} studies that identified proton-dominated energization \citep{Yin24b} and with in-situ observations showing stronger proton heating during reconnection in both the solar wind and Earth's magnetotail \citep{Phan13a,Ergun20,Oieroset23,Oieroset24,Rajhans25}. Although electrons undergo the same merger-driven Fermi process, their acceleration is limited by their lower initiation energy \citep{Yin25}.

These findings have broad implications for heliospheric and astrophysical plasmas in which reconnection occurs on macroscopic scales. In the heliospheric current sheet, Parker Solar Probe observations reveal signatures of multi-island reconnection and significant proton energization \citep{Desai25,Phan24}, consistent with the merger-driven dynamics explored here. In solar flares, the hierarchical formation and coalescence of magnetic islands  will naturally support the production of energetic ions whose energy content may exceed that of electrons \citep{Russell25}. Similar processes may shape particle acceleration in magnetotail reconnection \citep{Oieroset02,Ergun20b,Rajhans25}, reconnection-driven turbulence in the solar wind and large-scale reconnection in accretion flows and relativistic jets.

Overall, the present study provides the first quantitative determination of the maximum particle energy attainable in reconnection in macroscale systems and establishes the dynamics of island merging as the dominant factor controlling the impact of system size on maximum particle energy gain. This framework links microscopic energization physics to global system properties. 

These results also have implications for understanding the limitations of PIC and hybrid models in producing the extended powerlaw distributions of energetic particles seen in observations \citep{Li19,Zhang21,Zhang24}. The total number of mergers possible in a system is controlled by the range of flux ropes sizes that can develop in a simulation, with the caveat that the particles in these flux ropes must remain magnetized for efficient first order Fermi acceleration to occur. It is well known that protons demagnetize when island sizes fall below around 10$d_i$ with $d_i$ the ion inertial scale \citep{Mandt94}. Even in very large PIC simulations the largest islands are of order 40$d_i$ so the size ratio is only around four \citep{Zhang21}. This constraint severely limits the maximum particle energy gain. 

\begin{acknowledgments}
We acknowledge the support provided from NASA Grant Nos. 80NSSC20K1277, 80NSSC20K1813, 80NSSC24K0608 and
80NSSC22K0352, and NSF Grant No. PHY2109083. H.A. was supported by the NASA Early Career Investigator Program 80NSSC23K1061. The
simulations were carried out at the National Energy Research
Scientific Computing Center (NERSC). The data used to perform
the analysis and construct the figures for this paper are preserved at the NERSC High Performance Storage System and are available upon request.  
\end{acknowledgments}




\bibliography{sample7}{}
\bibliographystyle{aasjournalv7}

\end{CJK*}
\end{document}